\documentclass[aps,pre,showpacs,superscriptaddress,twocolumn,floatfix]{revtex4}
\usepackage{graphicx}
\usepackage{bm}
\usepackage{graphics}
\usepackage{amsmath}
\usepackage{amssymb}
\usepackage{wasysym}
\usepackage{color}

\def\be{\begin{equation}}
\def\ee{\end{equation}}
\def\bea{\begin{eqnarray}}
\def\eea{\end{eqnarray}}

\begin{document}

\title{Broad-tailed force distributions and velocity ordering in a
  heterogeneous membrane model for collective cell migration.}

\author{Tripti Bameta}
                                   
\affiliation{Department of Physics, Indian Institute of Technology, Bombay,
  Powai, Mumbai-400 076, India}

\author{Dipjyoti Das}
                                   
\affiliation{Department of Physics, Indian Institute of Technology, Bombay,
  Powai, Mumbai-400 076, India}

\author{Sumantra Sarkar}

\affiliation{Department of Physics, Indian Institute of Technology, Bombay,
  Powai, Mumbai-400 076, India}

\author{Dibyendu Das}
\email{dibyendu@phy.iitb.ac.in}
\affiliation{Department of Physics, Indian Institute of Technology,
  Bombay, Powai, Mumbai-400 076, India}

\author{Mandar M. Inamdar}
\email{minamdar@civil.iitb.ac.in}
\affiliation{Department of Civil Engineering, Indian Institute of Technology, Bombay, Powai, Mumbai-400 076, India}

\date{\today}

\begin{abstract}
Correlated velocity patterns and associated large length-scale
transmission of traction forces have been observed in collective live
cell migration as a response to a ``wound''. We argue that a simple
physical model of a force-driven heterogeneous elastic membrane
sliding over a viscous substrate can qualitatively explain a few
experimentally observed facts: (i) the growth of velocity ordering
which spreads from the wound boundary to the interior, (ii) the
exponential tails of the traction force distributions,
and (iii) the swirling pattern of velocities in the
  interior of the tissue.
\end{abstract}


\maketitle
\section{Introduction}
The phenomenon of collective cell migration arises in biological
processes of morphogenesis, wound healing, as well as cancer growth,
and is an active topic of current research interest \cite{review1,
  mark2010, dipole_collective, trepat, tambe, poujade1, poujade2}. To
understand the basic features in collective cell migration as a
response to wound healing, two-dimensional monolayer patches of
Madin-Darby canine kidney (MDCK) cells on deformable substrates have
been studied in different experiments
\cite{poujade1,poujade2,roure,trepat}.

For a physical scientist, there are many interesting aspects that
these experiments reveal. The spatially heterogeneous swarming and
swirling velocity patterns exhibited by the cells, studied by particle
image velocimetry~\cite{poujade1,poujade2}, are reminiscent of similar
pattern formation in active nematics and driven granular
matter~\cite{toner}.  As time passes, a zone of velocity order
starting from the wound boundary invades the interior of the MDCK
tissue \cite{poujade2}, reminding one of phase ordering kinetics
\cite{bray}. On the other hand, another set of experiments
\cite{roure,trepat} have shown that the local traction forces exerted
by the MDCK cells on the substrate have large fluctuations --- the
distribution of the forces being non-Gaussian with distinct broad
exponential tails, akin to force distributions in static granular
piles~\cite{Liu}. Yet, the MDCK cells forming the tissue are held to
each other and to the substrate by a network of Cadherin and Integrin
proteins, respectively~\cite{review1,alberts2008}, and they
self-generate active forces due to internal Actin and Myosin
dynamics. Thus, at the microscopic level, they show no resemblance to
mechanically driven loose granular rods or discs. Needless to say,
it is quite a challenge to model every observed feature of the MDCK
tissue system, as seen in different sets of experiments. In this
paper, we propose a simple statistical--mechanical model for the
system and show that in can simultaneously give a qualitative
explanation of the growth of velocity ordering with time, and the
large force fluctuations.

It is well known that the behaviour of large cell
collectives~\cite{earliest_collective} is qualitatively distinct from
that of a single cell~\cite{singlecell}.  Attempts have been made to
model cell assemblies incorporating signal transduction and cell-cell
signaling in Dictyostelium discoideum
\cite{cellsignal_collective}. Collective cell migration studies
incorporating cell division has been done \cite{cellgrowthkinetics1,
  cellgrowthkinetics2, arcerio2011, shraiman, maini}, but the
experiments that we are concerned with \cite{poujade1,poujade2,trepat}
have noted, that over the relevant time-scales, the growth of cell
number via cell division is not expected to play a role in the
features of interest in this paper. The geometrical instabilities such
as fingering and tip-splitting of the wound boundary in
experiments~\cite{poujade1,poujade2} have been theoretically modelled
using ideas of interface growth kinetics~\cite{front1,mark2010}.  The
velocity patterns in the interior of the
cell-sheet~\cite{poujade1,poujade2} have recently been studied by a
mechanical model, where the cells with a local orientation field
collectively behave in a viscoelastic
fashion~\cite{dipole_collective}. We show below that a much simpler
model compared to these can nevertheless capture two key features of
the MDCK tissue system, namely, the velocity ordering invasion, and
the large force fluctuations.  In particular, the novel aspect of our
model is to show the crucial role that heterogeneity plays in
determining the large force fluctuations.

The paper is arranged in the following fashion. In section~\ref{Model}
we describe our model in details, and justify its various components
on the basis of relevant experimental findings. We also specify the
equations and the parameter values used for the simulation
study of the model. In section~\ref{results}A--C we describe the three
major findings of our model. Numerical estimates to make quantitative
connections with the experiments are detailed in
section~\ref{num}. The significance of our model and findings in
comparison with previous theoretical works are discussed in
section~\ref{conc}. 
\begin{figure}
\includegraphics[scale = 0.3]{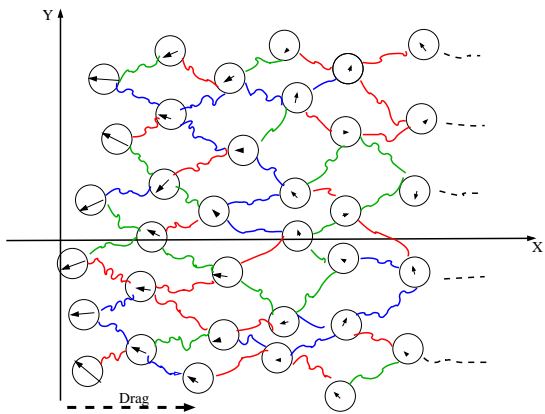}
\caption{A schematic picture (top view) of the deformed discretized
  membrane modeled as a tilted square lattice with undeformed
  spring-length $a_0$ of unity. The wound boundary cells move towards $-$ve
$X$. The cells are denoted by the circles and the active forces are
denoted by arrows (size proportional  the force magnitude) on
them. The varying spring stiffnesses are denoted by multiple
colors. The overall average substrate resistance is denoted by
  {\it Drag}.}
\end{figure}

\section{\label{Model}Model and Parameters}
We model the cell sheet as an elastic membrane,
  although some earlier studies have treated cell monolayers as
  viscoelastic~\cite{dipole_collective, review2_book}. The motivation
  for this choice comes from the experimental finding \cite{poujade2}
  that the average distance between MDCK cells does not change for
  many hours. It was clearly concluded that the movements of the cells
  are correlated showing very limited rearrangement, making the
  monolayer locally more elastic than viscous \cite{poujade2}.
  Similarly, a recent experimental study of internal stresses in cell
  monolayers \cite{tambe} have also assumed the latter to be elastic.
In our model, the cells are represented as discrete points in
continuous space, and the cell-cell cadherin connections are
represented by simple harmonic springs of effective stiffness
$\kappa_0$.  The elastic membrane is considered {\it heterogeneous}
with $\kappa_0$ drawn from a probability distribution function (p.d.f)
$P(\kappa_0)$.  Since there is inherent randomness in the number of
intercellular cadherin connections~\cite{Dobrowsky2008}, and the
possibility of any such connection to be in multiple distinct
structural states with different mechanical
properties~\cite{cadherin_exp1}, our assumption of heterogeneity seems
reasonable.  Next, it is expected that the connections of the cells
with the substrate via the integrin proteins will break and remake as
the tissue advances~\cite{alberts2008}. For a single cell, it has been
theoretically demonstrated that this complicated process of
cell-substrate interaction can be effectively replaced by a {\it
  linear} viscous drag \cite{const_sub_visco}. We extend the latter at
the tissue level and assume that the substrate exerts a local drag
force $-c_0 {\bf V}_i$ where ${\bf V}_i$ is velocity of the $i$-th
cell and $c_0$ is the drag constant.  The average position of the
``wound boundary'' (see Fig.~1) defines the $Y$-direction, and the
direction orthogonal to it will be called $X$. For simulating, we take
a $N \times N$ tilted square grid of points (Fig.~1), but while
thinking of a continuum limit, we will assume $N \rightarrow \infty$
such that effectively the wound boundary will be very far from the
center of the tissue (as in actual experiments
\cite{poujade2,trepat}).  Finally, we supply the ``live thrust
forces'' (originating from the cytoskeletal acto-myosin activity in
the cells) by hand in two alternate ways: 
(i) Cell$-i$ with position ${\bf R}_i = (X_i, Y_i)$ is given a space
and time dependent random force ${\bf F}_i = {\bf F}_{{\rm ave},i} +
\bm{\eta}_i$, with ${\bf F}_{{\rm ave},i} = -F_0 \exp(-n_{i}/\xi)
\hat{\bf x}$.  Here $n_i$ is the row number, from the boundary, of the
$i^{\rm th}$ cell. The boundary row is numbered $0$, and $\xi$ is a
length scale.  The components of ${\bm{\eta}}_i \equiv
(\eta_X,\eta_Y)_i$ are Gaussian white noise with zero mean and
$\langle \eta_{\alpha,i}(t_1)\eta_{\beta,j}(t_2) \rangle = 2\sigma
\exp(-n_{i}/\xi) \delta_{i,j} \delta_{\alpha,\beta} \delta(t_1 - t_2)$
with $\alpha$ and $\beta$ taking values $X,Y$. We will refer to this
as participatory model (PM) (see Fig.~1). Thus in the PM model, the
magnitude of the noise on force, just like the mean force, decays with
increasing distance from the boundary.
(ii) Only the cells at the wound boundary row are given a space and
time dependent random force with an average force ${\bf F}_{{\rm
    ave},i} = -F_0 \delta_{n_i,0} \hat{\bf x}$ added to Gaussian white
noise $\bm{\eta}_i$, with zero mean and $\langle
\eta_{\alpha,i}(t_1)\eta_{\beta,j}(t_2) \rangle = 2\sigma \delta_{i,j}
\delta_{\alpha,\beta} \delta(t_1 - t_2)$ ($\alpha \equiv \{X, Y\}$,
$\beta \equiv \{X,Y\}$).  We will refer to this as the leader driven
model (LDM).

The motivation for comparing
PM versus LDM comes from the discussions in Ref.~\cite{ladoux}
following the experiment of Ref.~\cite{trepat}.
The mathematical equation used to simulate the system is:
\begin{equation}
c_0 \frac{d {\bf R}_i}{d t} = \sum_{j} \kappa_{0}^{ij}(|{\bf R}_{j} -  {\bf R}_i| - a_0){\bf e}_{ij}  + {\bf F}_i.
\label{dyna1}
\end{equation}
The inertial term has been dropped 
  as the system is clearly overdamped. The
index $j$ in the sum in Eq. \ref{dyna1} goes over nearest neighbours
of $i$ and $\kappa_{0}^{ij}$ is the random stiffness constant of the
spring connecting $i$ and $j$. The unit vector ${\bf e}_{ij} = ({\bf
  R}_{j} - {\bf R}_i)/|{\bf R}_{j} - {\bf R}_i|$, and $a_0$ is the
undeformed length of the spring.
%
%

Here we report numerical results with parameter values: $c_0 = 10$,
$N=128$, $F_0 = 1$, $\sigma = 0.2$, $a_0=1$, and $P(\kappa_{0}^{ij})$
is a uniform box distribution between $0.25$ to $0.75$. 
We will show later that these choices of parameters
  lead to reasonable correspondence to experiments.  In the presence
of the active applied force ${\bf F}_i$, we time-evolve the positions
${\bf R}_i$ using Eq. \ref{dyna1}, and obtain the velocities ${\bf
  V}_i$ as $d{\bf R}_i/{d t}$. The simulation results for
Figs.~$2$, $3$, and $4$, are obtained by providing the cells with zero
initial velocities, and initial random displacements (from the
equilibrium positions) with components uniformly distributed over
$[-\delta,\delta]$. This may be expected to be a generic initial
condition for the cell collective. Periodic boundary conditions are
assumed along $Y$.

\section{\label{results}Results}
In this section we present the three important results of our model.

\subsection{Invasion of Velocity Ordering}

\begin{figure}
\includegraphics[scale=0.65,angle=0]{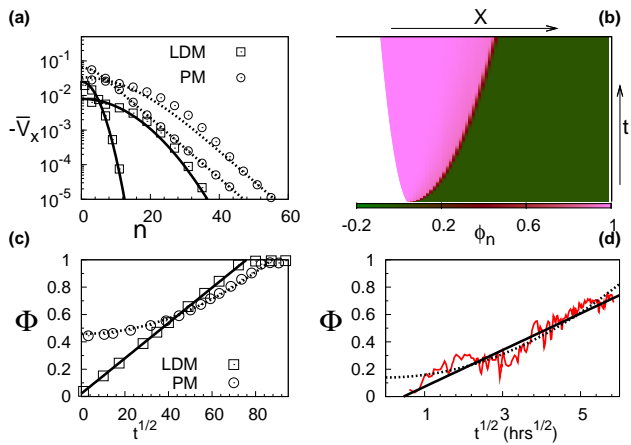}
\caption{Velocity ordering in the cell-sheet using $\delta =
  0.025$. (a) Velocity of the cell points as a function of row number
  for two different times ($100$ and $1000$).  The
  circles and squares represent simulation results for PM ($\xi = 5$)
  and LDM, respectively, and the fitting lines correspond to the
  theoretical model in Eq.~2. (b) The order parameter $\phi_n$ at
  spatial row position $x$ of the cell at different times
  $t$. The $x$-range is $0-120$ and $t$-range is
    $0-1000$. The velocity ordering front is clearly visible. (c) The
  order parameter $\Phi$ for the complete tissue, as a function of
  time $t$, with a fit of $t^{1/2}$ and $t$ for the LDM and PM,
  respectively. (d) Time dependence of experimental order parameter
  in~\cite{poujade2} fitted with $t$ and $t^{1/2}$.}
\end{figure}

We first proceed to show how under the action of active forces the
cell velocities acquire a bulk ordering. As the active forces are
preferentially oriented along $-\hat{\bf x}$ (for both the LDM and
PM), the velocities of the cells, starting with the ones close to the
boundary and followed by the ones in the bulk, gradually orient
themselves towards $-\hat{\bf x}$.  This velocity ordering is shown in
Fig. 2(a) --- with increasing time, for both the LDM and PM, mean row
velocity $-\overline{V_{X}}$ of deeper layers (with larger $n$) rise
in magnitude.  Thus an order-disorder boundary moves towards larger
$n$.  Interestingly, there is a quantitative difference between the
two models --- the shape of the $-\overline{V_{X}}(n)$ curve is
Gaussian for LDM and has an exponential tail for PM. A local order
parameter for every row $n$ can be defined as $\phi_n = - \overline{
  V_{{X},i}}/\overline{|{\bf V}_i|}$ (with all velocities $-V_{{X},i}
< 10^{-14}$ set to zero to avoid spurious contributions).  In
Fig. 2(b), $\phi_n$ is plotted (with larger magnitude corresponding to
lighter color) as a function of space $X$ and time $t$ for LDM --- the
movement of the order-disorder boundary with increasing $t$ is clearly
seen. A similar plot was observed experimentally in
\cite{poujade2}. Next, a global order parameter for the whole system
can be defined as $\Phi = \langle \phi_n \rangle_n$.  In Fig. 2(c),
$\Phi$ is shown to increase as $\sim t^{1/2}$ for LDM, and $\sim t$
for PM. To compare with the experiments, we have plotted the
experimental data for $\Phi$ from \cite{poujade2} in Fig. 2(d); two
curve-fits of $\sim t^{1/2}$ and $\sim t$ are put against the data,
showing that both these forms work reasonably well. Thus we have
demonstrated numerically that our models have similar velocity
ordering as seen experimentally in MDCK tissues \cite{poujade2} during
wound healing. We will now proceed to understand analytically which
ingredients of our model are essential for the above phenomenon, and
in particular the reason for quantitative differences between LDM and
PM.


It is interesting to note that the growth of $V_X(t)$ above can be understood 
analytically from an analogous $1$-dimensional problem. We can solve the
one-dimensional problem of a pulled, non-disordered, elastic chain:
\begin{equation}
c_0 \frac{\partial {u}}{\partial t} = \kappa_{0} \frac{\partial^2 {u}}{\partial x^2} + F(x,t).
\label{dyna2}
\end{equation}
The above equation is a simple $1$-D, linearized, continuum version of
Eq. \ref{dyna1}.  Here $u(x,t)$ is the displacement of any cell, $x
(\in [0,L])$ is the continuum space variable corresponding to the row
number $n_i$, and the non-random force is $F = -F_0 \exp(-x/\xi)$,
with $\partial u/\partial x = 0|_{x=0}$ (for PM), and $F = 0$, with
$\partial u/\partial x = F_0/\kappa_0|_{x=0}$ (for
LDM). The initial condition is taken as $u(x,0) = 0$ 
both for PM and LDM.
Eq.~\ref{dyna2} can be solved for velocity $v(x,t) = \partial
u/{\partial t}$ in the limit of large system size ($L
\rightarrow \infty$) with the boundary condition $u(\infty,t) = 0$. 
This gives unique analytical solutions:
\begin{eqnarray}
  {\rm PM}: v(x,t) - v_{\rm CM} & =&  -\frac{F_0}{2c_0} e^{(\tau - \tilde{x})}\Bigl( 1 + {\rm{Erf}}\bigl[\frac{\tilde{x} - 2 \tau}{2\sqrt{\tau}}\bigr] \nonumber \\
 & +& e^{2 \tilde{x}} ~{\rm Erfc}\bigl[\frac{\tilde{x} + 2 \tau}{2\sqrt{\tau}}\bigr] \Bigr),
\label{1dsolPM} \\
{\rm LDM}: v(x,t) - v_{\rm CM} & =& -\frac{F_0 }{\sqrt{c_0 \pi \kappa_0}} \frac{e^{-{c_0 x^2}/{4 \kappa_0 t}}}{\sqrt{t}}.
\label{1dsolLDM}
\end{eqnarray}
In Eq. \ref{1dsolPM} the symbols $\tilde{x} = x/\xi$ and $\tau =
\kappa_0 t/c_0 \xi^2$ are scaled dimensionless space and time
respectively, and Erf and Erfc refer to Error and Complementary Error
functions \cite{grad}, respectively. The profile of the ordered
velocity as a function of $x$ in the $1$-D LDM model
(Eq. \ref{1dsolLDM}) is clearly Gaussian, while in the PM model it is
modified (Eq. \ref{1dsolPM}) to have an exponential profile for large
$x$. The agreement in mathematical forms between the $1$-D analytical
result and the $2$-D simulation result in Fig.~2a shows that the
velocity ordering phenomenon is not particularly dependent on
stiffness randomness or noise, and its essence is captured even in a
$1$-D problem. As can be seen from Fig.~2c, the PM
  model gives a growth law $\Phi \sim t$ for a transient period, the
  reason for which may be understood from Eq.~3. For large
  ${\tilde{x}}$, ${\rm ln}(-v(x)) \sim {\tau} - {\tilde{x}} + {\rm
    Constant}$, to leading order, while for small ${\tilde{x}}$, ${\rm
    ln}(-v(x)) \sim {\tau} - {\tilde{x}^2}f({\tau}) + {\rm Constant}$
  (see PM in Fig.~2a),
  where $f({\tau})$ is a function of $ {\tau}$.  Thus there is an
  short distance quadratic profile, followed by a long distance linear
  profile in $x$. Spatial expanse of the quadratic profile keeps
  increasing, and beyond a certain time the growth law in PM will
  become $\sim t^{1/2}$ just like LDM.

We would now like to see if a drive from the boundary cell layer without any participation from the cells in the bulk can invoke other experimentally observed phenomena. Hence, in the rest of the paper we focus on LDM.
\subsection{Traction Force Fluctuations}

\begin{figure}
\includegraphics[scale=0.6]{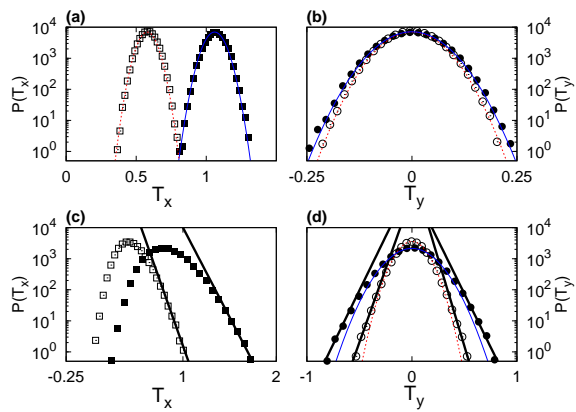}
\caption{Distributions $P({\bf T}_X)$ and $P({\bf T}_Y)$ of traction
  forces ${\bf T}_X$ and ${\bf T}_Y$ (in units of $10^{-3}$) for LDM,
  at $t = 300$ and $\delta = 0.025$.  Gaussian fits are shown for
  uniform $\kappa_0^{ij}$ in (a) and (b). For random $\kappa_0^{ij}$,
  deviation from Gaussianity, and resulting asymptotic exponential
  tails are shown in (c) and (d).  The data is for layers $18$ (filled
  symbols) and $19$ (empty symbols).}
\end{figure}

A second interesting result of our model is that the traction force
fluctuations are unusually large as in the experiments in
Ref.~\cite{trepat}.  In our model, the local effective traction force
on any cell is ${\bf T}_i = {\bf F}_i - c_0 {\bf V}$ (see
Eq. \ref{dyna1}). In LDM, ${\bf F}_i = 0$ (for any non-boundary cell)
and so ${\bf T}$'s are proportional to the local velocities ${\bf
  V}$. The probability distributions of the components ${\bf T}_X$ and
${\bf T}_Y$ for the LDM are shown in Fig.~3. For homogeneous membrane
(constant $\kappa_0^{ij}$), the distributions are clearly Gaussian
(Figs.~3(a--b)). To make sure that the latter is not a
  trivial consequence of the Gaussian distributed random thrust forces
  in the boundary layer, we checked the traction distributions when
  the boundary forces were drawn from a (i) box, and (ii) an
  exponential distribution. In both of these distinct cases, we found
  (data not shown) that the traction force components ${\bf T}_X$ and
  ${\bf T}_Y$ are Gaussian distributed. Thus there is no doubt that
  Central Limit Theorem (CLT) is valid and due to it, the local forces
  in the bulk (being sum of random neighbouring forces) turn out to be
  normally distributed. On the other hand, for a heterogeneous
membrane (random $\kappa_0^{ij}$) the distributions develop
exponential tails (Figs.~3(c--d)), indicating a departure from the
CLT. The shape of the curves of $P({\bf T}_X)$ (with mean at ${\bf
  T}_X \neq 0$) and $P({\bf T}_Y)$ (with mean at ${\bf T}_Y = 0$) have
qualitative resemblance to experiments --- in particular, the widths
decrease with increase of distance from the wound.


Breakdown of CLT in~\cite{trepat} is {\it a priori} quite
intriguing. It was speculated in Ref.~\cite{trepat}, that a $q$-model
\cite{Liu} like mechanism maybe at play.  Recognizing that the tissue
is heterogeneous, here we are specifically suggesting that an
effective $q$-model like mechanism may arise due to unequal stress
propagation and accumulation mediated by the random cadherin
connections.  Since the possibility of manipulating the strengths of
cadherin connections has been experimentally demonstrated
\cite{tambe}, our result is open to experimental test.

 The membrane can also be made heterogeneous in another way by
 introducing variable bond lengths of the cell-cell connections, while
 keeping the spring stiffness homogeneous. This naturally leads to
 deviation from regular lattice symmetry considered so far. The bond lengths 
$a_0^{ij}$ are made disordered by imparting new random equilibrium positions to the cells. To do so, the cells are shifted from the regular lattice by $\epsilon_x$ (along-X) and $\epsilon_y$ (along-Y), where $\epsilon_x$ and $\epsilon_y$ are drawn from uniform distribution over $\left[-\epsilon,\epsilon \right]$.
As can be seen from Eq.~1,
 the magnitude and direction of the spring force is independent of the
 bond-length (to the first order). Thus for ``small'' lattice
 distortion the distributions for $T_X$ and $T_Y$ are expected to be
 the same as that of the uniform non-distorted lattice. This is seen
 in our simulations for a choice of $\epsilon = 0.15$ ---the traction
 distributions are indeed Gaussian (see Figs.~4(a) and 4(b)).
 Contrary to this, with ``large'' random lattice distortion, one would
 expect from Eq.~1 random harmonic forces, leading to a departure from
 the latter result. We indeed see this when we make $\epsilon$ large
 (say $0.45$) ---the traction distributions develop exponential tails
 as shown in Figs. 4(c) and 4(d). Thus in two types of membrane
 heterogeneity ---random spring stiffnesses and random bond lengths
 --- we have found that non-Gaussian traction force distributions
 arise.

\begin{figure}
\includegraphics[scale=0.6]{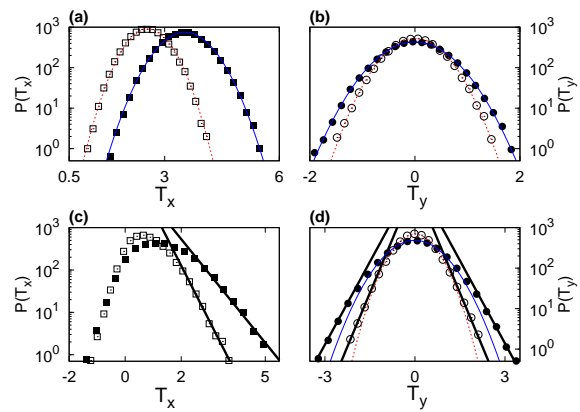}
\caption{Distributions $P({\bf T}_X)$ and $P({\bf T}_Y)$ of traction
  forces ${\bf T}_X$ and ${\bf T}_Y$ (in units of $10^{-3}$) for LDM,
  at $t = 300$, for a distorted lattice with random bond lengths. The uniform 
stiffness constant $k_0 = 1$. Gaussian fits are shown for $\epsilon = 0.15$ in (a) and (b). For
  $\epsilon = 0.45$, deviation from Gaussianity, and resulting
  asymptotic exponential tails are shown in (c) and (d).  The data is
  for layers $18$ (filled symbols) and $19$ (empty symbols).}
\end{figure}

\subsection{Swirling Patterns in the Bulk}
We now turn our attention to velocity patterns which
  develop in the bulk of the system.  The cells in the confined region
  of the experimental setup in refs.~\cite{poujade1, poujade2} are
  expected to be under internal stress due to the confinement from the
  boundary before the cell sheet is allowed to expand. In order to
  mimic this internal stress, we provide generic random initial
  positions to our cell lattice. The elastic relaxation of this
  ``pre-strained'' sheet help excite spatial modes through the
  harmonic couplings. The smaller wavelength modes damp out faster,
leaving large wavelength velocity swirls at late times (inset of
Fig.~5 for LDM). To precisely quantify the correlations in these
patterns, we show the velocity-velocity correlation function
$C_{vv}(\bar{n}) = \langle {\bf v}(n_0). {\bf v}(n_0+\bar{n})\rangle/
\langle {\bf v}^2 \rangle$ in Fig.~5; here $\bar{n} = y/{\sqrt{2}
  a_0}$ is the scaled distance in units of average inter-cellular
spacing along $Y$. The average $\langle \cdots \rangle$ is done over
ensembles as well as cell locations $n_0$, belonging to a strip in the
bulk region where $\langle {\bf v}(n_0) \rangle = 0$. We note that
$C_{vv}(\bar{n})$ shows change of sign beyond some layers (as seen
also in experiments \cite{weitzprl2010}) reflecting the bending of
velocity field over space. The correlation range increases with the
time of observation as expected --- it is $\approx 8-10$ cell layers
which is roughly similar as distances seen in experiments
\cite{poujade2,weitzprl2010}. The correlations
  observed in ref. \cite{weitzprl2010} are certainly influenced by
  substrate deformability and cell birth. But the fact that such
  correlations are also observed otherwise~\cite{poujade1,poujade2}
  indicates that they may not be the only factors governing the
  swirling patterns. Although our model cannot make these distinctions
  we note that it elucidates the role of inherent elasticity and
  pre-strain of the cell-sheet in producing such patterns.

\begin{figure}
\includegraphics[scale=0.5,angle=0]{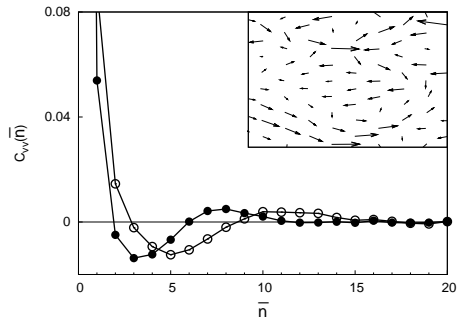}
\caption{The velocity-velocity correlation function $C_{vv}$ against
  scaled distance $\bar{n}$ along $Y$, at $t=500$ ($\bullet$) and
  $t=1000$ ($\circ$) for LDM using $\delta = 0.45$. Inset: Velocities
  of cells in the bulk region (dimension $8\times 8$)
  are shown (at $t=1000$) --- swirls can be clearly seen.}
\end{figure}

\subsection{\label{num}Numerical Estimates of Length, Time and Force Scales}

We have shown so far that a boundary layer driven heterogeneous
elastic sheet can produce qualitatively many experimental observations
in collectively migrating epithelial cells. We now make numerical
estimates of various quantities to find out whether our results are
quantitatively meaningful in comparison with the experiments.
 
 As shown in Fig. 1, the bond-length $a_0$ is unity. In real units it
 may be taken as $20 \mu {\rm m}$ (the average cell-cell
 separation~\cite{poujade2}).  To estimate the unit of time $t_0$, we
 start from Figs. 2(c) and (d).  For the LDM in 2(c), the slope of the
 line in units of $t_0^{-1/2}$ is $0.0139$, while the experimental
 data in 2(d) has a slope $0.139$ ${\rm hr}^{-1/2}$. Equating the two,
 gives $t_0 \approx 0.01 {\rm hr} = 36s$.  This tells us that the
 non-dimensional velocity in Fig. 2(a) in the boundary layer (for LDM)
 $\sim 10^{-2}$ corresponds to $10^{-2} a_0/t_0 \approx 20 \mu {\rm
   m/hr}$. This is in the ball-park of the velocities quoted in
 experiments \cite{poujade2}.  The curves in Fig. 2(a) are for times
 $t = 1$ hr ($100$ in simulation) and $10$ hrs ($1000$ in simulation).
 To make contact of forces in Fig. 3 with experiments, we choose to
 first obtain the bond-stiffness $\kappa_0$ in real units. The
 properties of a continuum cell sheet may be chosen as those appearing
 in the supplementary material of \cite{tambe}: Young's modulus $E =
 10$ kPa, Poisson's ratio $\nu = 0.5$, and sheet thickness of $h=5
 \mu$m.  We first consider a cell sheet of dimension $a_0 \times a_0
 \times h$, on which homogeneous tension $1$ Pa is applied. The
 relative change in surface area $\Delta A/A = 2(1-\nu)/E =
 10^{-4}$. On the other hand, applying equivalent edge force of $F=
 1{\rm Pa} \times a_0 \times h = 10^{-10}$ N on a $a_0 \times a_0$
 square, whose sides are made of springs of stiffness $\kappa_0$, we
 get $\Delta A/A = F/a_0 \kappa_0$.  Equating the two relative area
 changes, we get $\kappa_0 = 0.05$ N/m. But we have used an average
 stiffness value of $0.5$ in dimensionless units in our
 simulation. This gives the actual force unit to be $f_0 = 0.05 \times
 a_0/0.5$ N $= 2 \times 10^{-6}$ N.  In dimensionless units our forces
 in Fig. 3 for the $19^{\rm th}$ layer are $\sim 0.5 \times 10^{-3}$,
 which is equivalent to $ 10^{-9}$~N. This implies a traction of
 $10^{-9}/a_0^2 = 2.5$ Pa, which is one order of magnitude lower than
 that reported in \cite{trepat}. The dimensionless time $300$ reported
 in Fig. 3 translates to $\approx 3$ hrs. Using $a_0$, $t_0$ and
 $f_0$, we see that the dimensionless value of $c_0 = 10$ used in our
 simulation, is equivalent to $10 \times f_0 t_0/a_0 \approx 36$ N-s/m
 in real units. This can be converted into drag co-efficient $\zeta =
 c_0/a_0^2 = 25$ pN-hr/$\mu{\rm m}^3$, which is an order of magnitude
 lower than reported in \cite{dipole_collective}.  In Fig.~5, at our
 simulation time $t = 1000$ (equivalently $\approx 10$ hrs) the length
 scale associated with the first minimum of the correlation curve is
 $5 \times a_0 = 100 \mu$m. In \cite{weitzprl2010} at around $10$ hrs,
 a similar reported length scale is $200 \mu$m (or $\approx 10$ cell
 layers), which is higher than our result only by a factor of two.
Thus we see that, although our model is simple, it can make reasonably
close contact to experiment even quantitatively.  

One serious drawback is that our boundary tractions are very large
$\sim 10^3$ times compared to the bulk --- in reality \cite{trepat}
forces do not diminish so fast. It would be interesting to modify LDM
in the future by incorporating cellular thrusts from the bulk, to see
if the description becomes more realistic in a quantitative sense.

\section{\label{conc}Discussion and conclusion}

Recent experiments on collective migration of MDCK cells have
  thrown up several interesting puzzles, which in turn have spurred
  various theoretical modeling attempts.  Before we summarize the main
  results of this paper, we would like to situate our work with
  respect to the contributions made by the earlier theoretical
  models. The model of Ref.~\cite{dipole_collective} treats the cell
  sheet as a viscoelastic medium supplemented with a director field to
  describe the local cellular orientations. This model obtains the
  dependence of the velocity of the wound boundary on the viscoelastic
  parameters of the cell-sheet, and also shows complex correlated
  velocity patterns in the bulk. Another model~\cite{mark2010}
  concentrates specifically on the dynamics of the boundary of the
  cell-sheet. By introducing a competition between the curvature
  dependent driving force, and the elastic and viscous resistance of
  the cell sheet, the fingering instability as seen in
  experiment~\cite{poujade2} is reproduced by this model.  In contrast
  to these models, we treat the cell-sheet as an elastic membrane, and
  hope to capture some of the phenomena at early times. This is
  motivated by a direct experimental observation~\cite{poujade2}, and
  supported by treatment of cell-sheet as an elastic material in
  another set of experiments~\cite{tambe}.  We note that the phenomena
  we address in this paper, namely, growth of bulk velocity order
  parameter~\cite{poujade2}, and traction force
  distributions~\cite{trepat} have not been addressed in the
  aforementioned publications~\cite{mark2010, dipole_collective}. At
  the same time, the LDM cannot produce pronounced fingering due to
  the lack of {\it flowy} behaviour in our model.  A very recent
  paper~\cite{shraiman}, which studies the effect of cell
  proliferation and migration leading to contact inhibition,
  introduces a simple one-dimensional model, where the cells {\it
    plastically} spread in presence of cellular thrust forces from the
  boundary (similar to LDM in our paper). Nevertheless, since this
  model is one-dimensional, quite naturally, it cannot capture the
  two-dimensional phenomenology. 

In this paper we have identified few {\it minimal mechanical}
ingredients --- heterogeneous elastic membrane, fluid-viscous drag,
and the active drive of cells from the boundary to mechanically pull
the system --- which can explain three aspects of collective cell
migration: (a) macroscopic velocity ordering, (b) breakdown of CLT for
traction force fluctuations, and (c) velocity correlations associated
with swirls. Perhaps the most interesting result is, that membrane
heterogeneity has the capacity to induce broad tails in the
distribution of traction forces. The mechanism that we propose here is
very reminiscent of the q-model for static granular
assemblies~\cite{Liu}. At the same time, we would like to point out
that there are significant differences of our model from the
q-model. We have a mobile network of cells, velocity dependent
dissipative forces, and tensile force transmissions, as opposed to the
static transmission of compressive forces in the granular
assemblies. These differences may invite further analytical
exploration of our current model in the future.

The three results in the paper show a close qualitative resemblance to the
experiments. Even the quantitative estimates seem reasonable, albeit
with a major drawback that the values of traction forces and
velocities diminish much faster than the experimental values. This can
be attributed to the fact that our model does not pump energy in the
interior of the cell layer through active cellular thrusts. Recent
experiments~\cite{tambe} hint that cellular polarizations and cellular
active forces are possibly tied to mechanical stress cues from
surrounding cells. This demands our model to go beyond being purely
mechanical, by including a coupled dynamics (a cross-talk) between
active cellular forces and mechanical harmonic forces. While we would
explore these in future, we conclude by noting that this work sets a
benchmark by showing the achievements and limitations of a rather
simple mechanical model for collective cell migration.


\end{document}